\documentclass[acmsmall]{acmart}

\AtBeginDocument{%
  \providecommand\BibTeX{{%
    \normalfont B\kern-0.5em{\scshape i\kern-0.25em b}\kern-0.8em\TeX}}}

\setcopyright{rightsretained}
\copyrightyear{2020}

\acmYear{2020}
\acmConference[HATRA 2020]{HATRA 2020: Human Aspects of Types and Reasoning Assistants}{15--20 November, 2020}{Chicago, IL}
\acmBooktitle{HATRA 2020: Human Aspects of Types and Reasoning Assistants, 15--20 November, 2020, Chicago, IL}

\begin{document}

\title{Model-Driven Synthesis for Programming Tutors}

\author{Niek Mulleners}
\affiliation{%
  \institution{Utrecht University}
  \city{Utrecht}
  % \country{The Netherlands}
  }
\email{n.mulleners@uu.nl}
\author{Johan Jeuring}
\affiliation{%
  \institution{Utrecht University}
  \city{Utrecht}
  % \country{The Netherlands}
  }
\email{j.t.jeuring@uu.nl}
\author{Bastiaan Heeren}
\affiliation{%
  \institution{Open University of the Netherlands}
  \city{Heerlen}
  % \country{The Netherlands}
  }
\email{bastiaan.heeren@ou.nl}

\begin{abstract}

When giving automated feedback to a student working on a beginner's exercise, many programming tutors run into a completeness problem.
On the one hand, we want a student to experiment freely.
On the other hand, we want a student to write her program in such a way that we can provide constructive feedback.
We propose to investigate how we can overcome this problem by using program synthesis, which we use to generate correct solutions that closely match a student program,
and give feedback based on the results.

\end{abstract}




\maketitle

\section{Introduction}

Programming tutors help students learning to develop a program.
Since the 1960s, more than one hundred of such tools have been developed~\citep{keuning_2018}.
Many of these tutors analyse the final program a student submits, but there are also tutors that analyse a partial, unfinished student program, and give feedback for such programs.
\citet{keuning_2018} note that most of these tools give solution error feedback and that relatively few tools give hints to students when they are on their way to a solution.
We want to investigate how we can give hints to students that are developing a program, at a stage where a program need not be complete yet.

Giving hints to students who are halfway towards a solution is challenging.
The literature describes several techniques to achieve this.
\citet{anderson_1986} and \citet{gerdes_2017} require a student to follow a path towards a model solution, where there are usually multiple paths possible.
\citet{price_2016}, \citet{piech_2015}, and \citet{lazar_2014} use data collected from previous students to generate hints.
The disadvantages of these approaches are that they either restrict what a student can do, or that they are incomplete in the sense that a student can submit a program for which no hint can be given.
The latter need not be a problem if this happens infrequently, but current practice shows that this happens regularly.
We want to investigate techniques for giving hints to students who can develop any program for a problem, and ask for hints at any stage.

We will focus on students working on beginners exercises in a statically typed functional programming language.
Such an exercise is defined by

\begin{itemize}
  \item a textual description;
  \item a set of model solutions;
  \item a set of properties;
  \item a library of allowed functions.
\end{itemize}

We allow students to leave parts of their program unimplemented by introducing holes, denoted by a question mark (\verb|?|).
For an unfinished student program, we want to answer the following questions:
is the student still on the right track?
If so, how can we give hints on how to proceed?
If not, how can we help the student to get back on track?

\citet{feldman_2019} use program synthesis in a dynamically typed functional language to tell students whether they are on the right track,
by generating the part of the program the student still needs to write.
\citet{rishabh_2013} and \citet{gulwani_2018} apply program synthesis to imperative languages to give automated feedback on how to repair a program.
We want to adapt these approaches to a statically typed functional language and use the synthesis results to give constructive feedback. 
We propose to investigate to what extent we can use program synthesis to give hints to students on their way to a solution.

\section{An example: sorting algorithm}

Throughout this proposal, we use the following exercise as an example.

\subsection*{Description}

Write a function \verb|my_sort :: [Int] -> [Int]| that sorts a list of integers.

\subsection*{Model solutions}

There are many possible ways to implement a sorting algorithm.
For simplicity, we only look at the following model solution, implementing \verb|my_sort| as insertion sort.

\begin{verbatim}
my_sort = foldr insert []
  where 
    insert x [] = [x]
    insert x (y:ys) | x < y     = x:y:ys
                    | otherwise = y:insert x ys
\end{verbatim}

\subsection*{Properties}

The correctness of \verb|my_sort| is expressed in terms of the following properties,
expressing that the result of sorting a list should be a permutation of the input list and that the result should be in non-descending order.

\begin{verbatim}
sort_permutes xs = my_sort xs `permutes` xs 
sort_nondescending xs = nondescending (my_sort xs)
\end{verbatim}

\subsection*{Library}

A suitable prelude containing all functions the students should reasonably be allowed to use.

\section{Program Synthesis}

Program synthesis is the process of automatically generating programs based on their intended behaviour.
Rather than writing programs by hand, the programmer expresses their intention in terms of a specification to which the program should adhere.
Such a specification can range from simple input-output examples to a formal logical specification.
A program synthesizer then tries to generate a program from this specification, typically by performing a search, 
enumerating all possible programs within the program space, until one is found that adheres to the specification.

To verify the correctness of student solutions, we extensively test them against a large set of input-output examples, generated from the model solutions.
Therefore, as our synthesizer, we will look at \textsc{Myth} \citep{osera_2015,frankle_2016}, a state-of-the-art tool in example-based synthesizers.
During the synthesis process, \textsc{Myth} can only introduce constructs that it can execute, 
so that the synthesized program can be checked against the input-output examples.
When synthesizing a recursive function, \textsc{Myth} can ``execute'' a recursive call if the input to the recursive call is in the example set,
by simply returning the corresponding output.
This requires, however, that the example set is \textit{trace complete}.
That is, for every input-output pair in the example set, there should also be an input-output pair for every structurally smaller input, 
so that the example set contains complete program traces.
This requirement is often considered the main limitation of \textsc{Myth}.
In the presence of model solutions, however, we can simply execute a model solution at the recursive call, bypassing the need for trace-complete example sets! 

Furthermore, we expect correct solutions to resemble our model solutions.
We can use this insight to guide the synthesis process by giving priority to constructs appearing in the model solutions.
\citet{lee_2018} accelerate search-based synthesis by performing A* search \citep{hart_1968}, with a heuristic based on learned probabilistic models.
We intend to use a similar approach to speed up the synthesis, by basing our heuristic on the provided model solutions.

Unfortunately, \textsc{Myth} cannot be applied directly to our setting, because it requires local input-output examples on holes,
whereas we only have global input-output examples.
As described by \citet{lubin_2020}, we can use live bidirectional evaluation to propagate global input-output examples to local input-output examples on holes.
First, live evaluation~\citep{omar_2019} is used to partially evaluate a program to a result \textit{r} by proceeding around the holes.
Live unevaluation~\citep{lubin_2020} then checks this result \textit{r} against an input-output example, 
generating constraints on the holes in the form of local input-output examples. 

\section{Finishing programs}

In this section, we give an overview of how we intend to apply the described techniques to complete unfinished student programs and use the results to give feedback.

\subsection{Live bidirectional example checking}

Before we can perform program synthesis, we apply live bidirectional example checking to check the student solution against an example set, 
possibly returning a set of local input-output examples for every hole.

\subsubsection{Failure}

If live bidirectional example checking fails, this means that it is impossible to finish a student solution by filling its holes.
This results in a counterexample, which can be checked against the property set to determine which properties are violated.

\paragraph{Finished solution}

The student might try to implement \verb|my_sort| as insertion sort, but incorrectly use the list cons operator instead of the \verb|insert| function.

\begin{verbatim}
my_sort = foldr (:) []
\end{verbatim}

Live evaluation might return the counterexample \verb|my_sort [3,2,1] == [3,2,1]|, violating the property \verb|sort_nondescending|.

\paragraph{Partial solution}

Alternatively, the student might give a partial solution with an incorrect base case, such as the following.

\begin{verbatim}
my_sort = foldr ? [0]
\end{verbatim}

Live evaluation gives the counterexample \verb|my_sort [] == [0]|, which violates the property \verb|sort_permutes|.
Note that, in some cases, the resulting counterexample might also contain a hole!
For example, take the following partial solution. 

\begin{verbatim}
my_sort [] = []
my_sort (x:xs) = x : ?
\end{verbatim}

Live evaluation might return the counterexample \verb|my_sort [2,1] == 2:?|.
Using unevaluation, we can determine that there is no hole filling for which the epxression \verb|2:?| evaluates to the expected result \verb|[1,2]|. 

\paragraph{Recovery} \label{recovery}

Other than presenting the student the counterexample and showing which properties her implementation violates, we might want to help her get back on track.
Each counterexample corresponds to a path in the abstract syntax tree of the program, hinting at what part of the program is incorrect.
For example, the counterexample \verb|my_sort [3,2,1] == [3,2,1]| is introduced by the leaf \verb|(:)| in the abstract syntax tree of \verb|foldr (:) []|.
We replace this leaf with a hole (resulting in the program \verb|foldr ? []|) and repeat the synthesis process until we eventually find a correct solution closest to the student's attempt.

\subsubsection{Success}

On successful live bidirectional evaluation, we get a set of local input-output examples on the holes.
If the student solution contains no holes, this set of examples is empty and the exercise is finished.
If there are holes, the generated examples can be used to perform program synthesis.

\subsection{Program synthesis}

Now that we have local input-output examples on the holes in our program, we can perform program synthesis to determine whether the student is still on track. 

\subsubsection*{Failure}

If a program can no longer be finished, we expect our synthesis to fail.
This means that the synthesis has either exhaustively searched the program space or a timeout occurred.
Both of these cases can take quite some time.
One way to skip this process is to check for conflicting examples.
Take, for example, the following attempt at implementing \verb|my_sort| using \verb|map|. 

\begin{verbatim}
my_sort = map ?0
\end{verbatim}

For the input \verb|[2,2,1]| we expect the output \verb|[1,2,2]|.
As such, for a function \verb|f| filling the hole we generate the following input-output examples.

\begin{verbatim}
f 2 == 2
f 2 == 1
f 1 == 2  
\end{verbatim}

These constraints are conflicting, since a function cannot have different outputs on the same input.
As such, we can skip the synthesis process.
For other incorrect programs, it might not be possible to generate conflicting constraints.
Consider the following example:

\begin{verbatim}
my_sort = map ? . zip [0..]
\end{verbatim}

Before mapping, it first combines every value in the input list with a unique integer.
In doing so, the list passed to \verb|map ?| contains only unique values, so that no conflicting constraints can be generated.
At this point, we will perform program synthesis with the generated constraints,
but the process will fail since \verb|map| cannot change the order of a list.

\paragraph{Recovery}

To help the student recover, we propose an approach similar to the one described in section \ref{recovery}.
If the synthesis process fails to fill a hole, we blame its parent node in the abstract syntax tree of the program and replace that node with a hole.
For example, for the incorrect program \verb|map ? . zip [0..]|, we replace \verb|map ?| with a hole, resulting in the program \verb|? . zip [0..]|.
Like before, we repeat the synthesis process until we find a correct solution.

\subsubsection*{Success}

If a program can still be finished, our synthesis will either timeout, or it will return a valid hole filling.
In the first case, we cannot tell a student whether she is on the right track,
but we can guarantee that her implementation is more complex than required and advise another approach.
In the second case, we can give her feedback by using the synthesized programs to calculate specifications for the holes and present these in a legible way.
For example, take the following partial student program.

\begin{verbatim}
my_sort [] = []
my_sort (x:xs) = f x (my_sort xs)
  where f y ys = ?
\end{verbatim}

This program can easily be finished by filling the hole with \verb|insert y ys|.
Knowing this, we can generate a specification for the hole, by generating input-output examples using the hole filling.
These input-output examples can help the student understand the part of the program they are yet to write.
Even more interesting, we might use a technique such as described by \citet{claessen_2010} to automatically generate properties that should hold on \verb|insert| and therefore should hold on the hole.
Ideally, this would allow us to inform the student that \verb|ys| is a sorted list and \verb|f| should return a sorted list containing all elements from \verb|ys| plus \verb|y|.

Note that many existing techniques, such as program transformations \cite{gerdes_2017},
can already give relevant feedback on the previous example by comparing it to one of the model solutions.
The next example, however, is a lot harder to reason about automatically.

\begin{verbatim}
my_sort [] = []
my_sort (x:xs) = foldr ? ? xs
\end{verbatim}

It is far from trivial to automatically prove that this partial program is still on the right track,
but we expect our synthesizer to find the correct hole fillings (\verb|insert| and \verb|[x]| respectively) with relative ease.

\section{Conclusion}

To conclude, we believe there is a lot of potential in the application of program synthesis and live bidirectional evaluation
to give meaningful feedback to students on their way to a solution.

\bibliographystyle{ACM-Reference-Format}
\bibliography{citations}

%\appendix

\end{document}